\documentclass[11pt]{article}
\usepackage{fullpage}
\usepackage{psfig}

\newtheorem{theorem}{Theorem}
\newtheorem{definition}{Definition}

\newtheorem{lemma}[theorem]{Lemma}

\newcommand{\qed}{\mbox{\ \ \ }\rule{6pt}{7pt} \medskip}
\newcommand{\oz}{0/1}
\newcommand{\calt}{{\cal{T}}}
\newcommand{\cals}{{\cal{S}}}
\newcommand{\calb}{{\cal{B}}}
\newcommand{\calf}{{\cal{F}}}
\newcommand{\no}{{\sc{NO}~~}}
\newcommand{\sat}{{\sc{SAT}~~}}
\newcommand{\yes}{{\sc{YES}~~}}

\newcommand{\etal}{{\em et al.}} 
\newcommand{\ra}{\rightarrow}
\newcommand{\la}{\leftarrow}
\newcommand{\lra}{\leftrightarrow}
\def\proofof#1{\paragraph{Proof of #1.}}

\newenvironment{proof}{\noindent{\em Proof:~}}{\hfill\qed}

\title{{The Complexity of Quantum Systems on a One-dimensional Chain} }
\author{
Sandy Irani\thanks{ Information and Computer Science
Department, UC Irvine, 92697, {\tt irani@ics.uci.edu}. Partially
supported by NSF Grant CCR-0514082. } }
\date{\today}

\makeatletter

\begin{document}

\maketitle 

\begin{abstract}
We prove that adiabatic computation is equivalent to standard quantum computation
even when the adiabatic quantum system is restricted to be a set of particles
on a one-dimensional chain. We give a construction that uses a 2-local Hamiltonian
on nearest neighbors using particles that can have
ten distinct states. This implies a construction
of a one-dimensional chain of qubits in which the Hamiltonian is 6-local.
We adapt this construction to show that the 2-local Hamiltonian for
$13$-state particles is QMA-complete which in turn implies that
the 8-local Hamiltonian 
restricted to a one-dimensional
chain of qubits is QMA-complete.
\footnote{Similar results have also been obtained by Aharonov, Gottesman and
Kempe \cite{AGK06}.}
\end{abstract}

\section{Introduction}

Adiabatic computation was introduced by Farhi, Goldstone, Gutmann and Sipser
\cite{farhi-2000} as a means
of solving difficult classical optimization problems.
Although it appears likely that algorithms for this model
require exponential time to solve NP-hard problems, 
adiabatic computation remains an appealing alternative to standard computation
since it may be more robust against certain types of quantum errors \cite{childs-2002-65}.

The idea behind adiabatic computation is to define two Hamiltonians
$H_{init}$ and $H_{final}$. The ground state of $H_{init}$ 
(the eigenstate coresponding to the smallest eigenvalue)
should be easy to prepare, like
a tensor product states. The desired output is the ground state of $H_{final}$
which should somehow encode the solution to the problem.
We also require that both Hamiltonians be local in that they are the sums of
Hamiltonians, each of which operates on a constant number of particles.
Thus, they are efficiently described by enumerating the matrices associated
with each component. 
The Hamiltonian is slowly varied from $H_{init}$ to $H_{final}$
by increasing $s$ from $0$ to $1$ in the following expression:
$H(s) = (1-s) H_{init} + s H_{final}$.
The adiabatic theorem (expressed more formally in the next section)
states that if a system begins in the ground state of $H_{init}$ and
the system is varied slowly enough, it will remain in the ground state.
Thus, the final state will be the desired ground state of $H_{final}$.
The required time to vary $s$ from $0$ to $1$ depends on 
the minimum spectral gap of $H(s)$.
In order to think of adiabatic quantum computation as computing a 
classical function (as opposed to quantum states), a measurement of
one or more of the particles is then performed to yield a 
classical output.

van Dam, Mosca and Vazirani proved  that the standard model of quantum computation is at least
as strong as the adiabatic quantum model \cite{vandam-2002}.
Subsequently, Aharonov \etal\ showed that an arbitrary quantum circuit can be simulated 
by an adiabatic computation \cite{ADKLLR04}, thus establishing polynomial
equivalence of the two models. 
Furthermore, the construction in \cite{ADKLLR04} showed that any quantum computation
could be efficiently simulated by an adiabatic computation with
$2$-local nearest-neighbor Hamiltonians operating on six-state particles
on a two dimensional grid. This was later improved by Oliveira and Tehral
\cite{oliveira-2005}
to a two-dimensional grid of qubits (two state particles).
We prove the following theorem for 1-dimensional systems:
\begin{theorem}
An arbitrary quantum circuit can be simulated 
by an adiabatic computation with $2$-local nearest neighbor
Hamiltonians operating on a 1-dimensional chain of ten-state particles.
This in turn implies that the construction will work on a 1-dimensional chain
of quibits with $6$-local Hamiltonians, each of which operates on continguious
sets of qubits. 
\end{theorem}

A slight modification of our
construction has implications for the study of the complexity of
quantum systems. 
The study of quantum systems from the point of view of 
their computational complexity is an important part of the more general
research program aimed at understanding the power of quantum computing.
In particular, many one-dimensional quantum systems appear to 
be more tractable in some ways than their two-dimensional 
counterparts \cite{vidal-2004-93}.
This paper seeks to answer whether certain hardness results for
two-dimensional quantum systems can be carried over to the one-dimensional
case. In particular, can a one-dimensional quantum system have a sufficiently
rich structure to implement universal computation?

Proving that a given question about a quantum system is QMA-complete provides
strong evidence that it is computationally difficult, even with the
benefit of a quantum computer.
The class QMA is the quantum analog of NP and MA. 
That is, it is the set of all languages that
can be probabilistically verified by a quantum verifier in polynomial
time. Kitaev initiated the study of QMA-complete problems by defining
the local hamiltonian problem, the quantum analog of \sat \cite{863284}.
In this problem, one is given a Hamiltonian and a guarantee that the
lowest eigenvalue for $H$ is either greater 
than some value $b$ or
less than another values $a$, where $b-a$ is at least an inverse polynomial
in the number of qubits in the system. The output of the problem is
to determine which alternative is the case for $H$.
Kitaev gave the first  QMA-complete problem by showing that the $5$-local
Hamiltonian is QMA-complete \cite{863284}.
This was later improved by Kempe and Regev who showed that the $3$-local Hamiltonian
is QMA-complete \cite{kempe-2003-3}
and then by Kempe, Kitaev and Regev who showed that the $2$-local Hamiltonian
is QMA-complete \cite{kempe-2006-35}. 
Finally Oliveira and Tehral \cite{oliveira-2005} 
showed that this latter result holds even when
the $2$-local Hamiltonian is constrained to be nearest-neighor interactions
on a grid of qubits. They conjecture that it is not possible to extend this
result for a $1$-dimensional chain. We give the following result:
\begin{theorem}
The $2$-local Hamiltonian on a $1$-dimensional chain of
$13$-state particles is QMA-complete. 
This in turn implies that the $8$-local Hamiltonian on a $1$-dimensional chain
is QMA-complete. 
\end{theorem}
Certainly an intriguing question left open here is
whether the locality can be reduced and, in particular, 
whether the $2$-local Hamiltonian on a $1$-dimensional chain
is QMA-complete.

Similar results to these have been
independently obtained by Aharonov, Gottesman and Kempe \cite{AGK06}.
\section{Preliminaries}

\subsection{The Model of Adiabatic Computation}
\label{subsection:adiabatic}

The Adiabatic Theorem, stated here, is the
foundation of adiabatic computation.
\begin{theorem}
\label{th:adiabatic}
{\bf The Adiabatic Theorem (as adapted from \cite{Reichardt04} and quoted from \cite{ADKLLR04} )}
Let $H_{init}$ and $H_{final}$ be two Hamiltonians acting on a quantum system and consider
the time-dependent Hamiltonian $H(s) = (1-s) H_{init} + s H_{final}$.
Assume that for all $s$, $H(s)$ has a unique ground state.
Then for any fixed $\delta > 0$, if
\begin{equation}
\label{eq:lb}
T \ge \Omega \left(
\frac{ \| H_{final} - H_{init} \|^{1 + \delta} }
{\epsilon^{\delta} \min_{s \in [0,1]} \{
\Delta^{2 + \delta} ( H(s))
\} }
\right) ,
\end{equation}
 the final state of an adiabatic evolution according to $H$ for time $T$
(with an appropriate setting of global phase) is
$\epsilon$-close in $l_2$-norm  to the ground state of $H_{final}$.
The matrix norm is the spectral norm
$\| H \| = \max_w \| H w \| / \| w \|$.
\end{theorem}
$\delta$ will be a contstant and the constant in the $\Omega$ will 
go to infinity as $\delta$ goes to $0$.
We use the model of adiabatic computation as described in
\cite{ADKLLR04}:

\begin{definition}
A $k$-local adiabatic computation $AC(n, d, H_{init}, H_{final}, \epsilon )$
is specified by two k-local Hamiltonians, $H_{init}$ and $H_{final}$
acting on $n$ $d$-dimensional particles, such that both Hamiltonians
have unique ground states. The ground state of $H_{init}$ is a tensor
product state.
The output is a state that is $\epsilon$-close in $l_2$-norm to the ground
state of $H_{final}$. Let $T$ be the smallest time such that the final
state of an adiabatic evolution according to $H(s) = (1-s) H_{init} + s H_{final}$
for time $T$ is $\epsilon$-close in $l_s$-norm to the ground state of
$H_{final}$. The running time of the adiabatic algorithm is
defined to be $T \cdot \max_s \| H(s) \|$.
\end{definition}
Thus, the running time of the computation will be upper bounded by the norm
of the Hamiltonians and the lower bound given in Inequality \ref{eq:lb}.

It is possible in some cases to restrict our attention to a subspace $\cals$.
Suppose that $H(s)$ leaves $\cals$ invariant.
$H_{\cals}(s)$ is the restriction of $H(s)$ to $\cals$.
If we start the adiabatic evolution at a state inside $\cals$
then an adiabatic evolution according to $H_{\cals} (s)$ is identical
to an adiabatic evolution according to $H(s)$.
Therefore, we can use $\Delta H_{\cals} (s)$ in Theorem
\ref{th:adiabatic} instead of $\Delta H(s)$.

\subsection{The Class QMA and the Local Hamiltonian Problem}

The class QMA is defined in terms of promise problems defined by
a pair $(L_{yes},L_{no})$ of disjoint sets of strings corresponding
to \yes and \no instances of the problem. 
The input is guaranteed to be in $L_{yes} \cup L_{no}$ and
solving the problem entails determining whether a given input
string $x$ is in $L_{no}$ or $L_{yes}$.
$\calb$ is defined to be the Hilbert space of a single
quibit.

\begin{definition}
{\bf (QMA)}
Fix $\epsilon = \epsilon( | x|)$ such that $\epsilon = 2^{-\Omega(|x|)}$.
Then, a promise problem $L \in QMA$ if there exists a quantum polynomial
time verifier $V$ and a polynomial $p$ such that
$$ \forall x \in L_{yes}, \exists | \xi \rangle \in
\calb^{\otimes p(|x|)}, Pr(V( | x \rangle, | \xi \rangle) = 1) \ge 1 -\epsilon$$
$$ \forall x \in L_{no}, \forall  | \xi \rangle \in
\calb^{\otimes p(|x|)}, Pr(V( | x \rangle, | \xi \rangle) = 1) \le \epsilon$$
where $Pr(V( | x \rangle, | \xi \rangle) = 1)$ denotes the probability that $V$
outputs $1$ given $| x \rangle$ and $| \xi \rangle$.
\end{definition}

The first known QMA-complete problem is  the local Hamiltonian problem
and is a natural analog of \sat.

\begin{definition}
We say that an operator $H: \calb^{\otimes n} \rightarrow \calb^{\otimes n} $
on $n$ qubits is $k$-local if H can be expressed as the sum of terms,
where each term is a Hermitian operator acting on at most $k$ bits.
\end{definition}

\begin{definition}
{\bf (The (promise) problem k-local Hamiltonian )}
A $k$-local Hamiltonian on $n$ qubits $H = \Sigma_{j=1}^r H_j$ with $r = poly(n)$.
Each $H_j$ has a bounded operator norm $ \| H_j \| \le poly(n)$
and its entries are specified by $poly(n)$ bits.
In addition, we are given two constants $a$ and $b$ with $a<b$.
In \yes instances, the smallest eigenvalue of $H$ is at most $a$.
In \no instances, it is larger than $b$.
\end{definition}

\subsection{Previous Results on Spectral Gaps}
\label{sec:SpectralGaps}

We make use of some previous results on spectral gaps which we
state here in a form that is
particularly suited for our purposes.
Define $P_r$ to be the $r \times r$ matrix of the
following form:
\[
\left(
\begin{array}{rrrrrrr}
 \frac 1 2  &  -  \frac 1 2  & 0 & & \cdots & & 0\\
 -\frac 1 2 & 1   &  - \frac 1 2  & 0  & \ddots & & 0\\
0 &  -\frac 1 2 & 1   &   -\frac 1 2  & 0  & \ddots &  \vdots \\
 & \ddots & \ddots & \ddots & \ddots  & \ddots &   \\
\vdots & & 0 &  -\frac 1 2  & 1 &  -\frac 1 2  & 0\\
0 & & \cdots & & 0 &  -\frac 1 2  &  \frac 1 2 
\end{array} 
\right)
\]

\begin{lemma}
\label{lem:gap}
Consider a Hamiltonian $H = H_1 + H_2$ acting on a subspace $\cals$.
Consider a basis for $\cals$, $\gamma_1, \ldots, \gamma_r$.
Suppose that when $H_1$ is restricted to $\cals$ and expressed in
this basis, the matrix is $P_r$.
Suppose also that when $H_2$ is restricted to $\cals$ and expressed 
in this basis, the matrix is diagonal with non-negative integer entries,
at least one of which is nonzero.
Then the lowest eigenvalue of $H$ when restricted to $\cals$ is
$\Omega( 1 / r^4)$.
\end{lemma}

For the proof of this lemma, we make use of the following lemma proven
by Kitaev (Lemma 14.4 in \cite{863284}).

\begin{lemma}
\label{lem:geom}
Let $H_1$ and $H_2$ be two Hamiltonians with ground energies $a_1$ and $a_2$,
respectively. Suppose that for both Hamiltonians the difference between the energy
of the (possibly degenerate) ground space and the next highest eigenvalue is larger
than $\Lambda$, and that the angle between the two ground spaces is $\theta$.
Then the ground energy of $H_1$ + $H_2$ is at least
$a_1 + a_2 + 2 \Lambda \sin^2(\theta / 2 )$.
\end{lemma}

\proofof{~~Lemma \ref{lem:gap}}
Suppose that the diagonal entries are all non-zero. Since the entries
are integral, the lowest eigenvalue of $H_2$ is at least $1$.
Since $H_1$ and $H_2$ are both positive semi-definite,
the ground energy of $H_1 + H_2$ is at least 1.
Now suppose that $H_2$ has at least one non-zero entry. It's spectral gap
is at least $1$ since it has at least one non-zero entry on its diagonal.
Using standard techniques, one can show that the spectral gap of $H_1$
is at least $\Omega( 1 / r^2 )$. The ground energies of $H_1$ and $H_2$
are both $0$. Furthermore since the ground state of $H_1$ is
a uniform superposition of all the basis vectors and $H_2$ has at least
one non-zero entry, the cosine of the angle between
 the two ground spaces is at most $1 - 1/r$.
 Invoking Lemma \ref{lem:geom}, we have that
the ground energy of $H_1 + H_2$ is at least $\Omega( 1 / r^3)$.
\qed

The next lemma is instrumental in providing a spectral gap for the
adiabatic computation:

\begin{lemma}
\label{lem:sgap}
Let $H_{final}$ be $P_r$ for some $r$.
Let $H_{init}$ be an $r \times r$ matrix of the following form:
\[
\left(
\begin{array}{rrrr}
0 & 0 & \cdots & 0\\
0 & 1 & \cdots & 0 \\
\vdots & \vdots & \ddots & \vdots \\
0 & 0 & \cdots & 1\\
\end{array} 
\right)
\]
Define $H(s) = (1-s) H_{init} + s H_{final}$.
Then the spectral gap of $H(s)$ is $\Omega(r^{-2})$
for all $s \in [0,1]$.
\end{lemma}
This lemma is proven in \cite{ADKLLR04}.

\section{A One-dimensional Universal Quantum Adiabatic Computer}

The proof of the following theorem shows that we can simulate a quantum
circuit that uses $L$ gates on $n$ qubits with
an adiabatic computation on a chain of $(L+2)$ ten-state
particles on a line. 

\begin{theorem}
Given a quantum circuit on n qubits with L two-qubit gates
implementing a unitary U, and $\epsilon > 0$, there is a
$2$-local adiabatic computation between nearest-neighbor
particles on a line
$AC(L+2, 10, H_{init}, H_{final}, \epsilon )$
whose running time is polynomial in $L$ and $1/\epsilon$
and whose output is $\epsilon$ close to $U | 0^n \rangle$.
Moreover, $H_{init}$ and $H_{final}$ can be computed by a polynomial
time Turing machine.
\end{theorem}

We first show a consturction that uses particles with thirteen states.
We will  add one extra state at the very end and then reduce the
number of states by four by
 identifying
 pairs of states. The final construction will use ten states.
 Our proof makes use of the observation in 
Section \ref{subsection:adiabatic} that one need only prove a lower
bound on the spectral gap of $H(s)$ when restricted to a subspace
as long as the adiabatic computation starts in that same subspace.

We follow the convention in \cite{ADKLLR04} which assumes without
loss of generality that the quantum circuit to be simulated has a particular
layout of the gates. It consists of a sequence of $R$ rounds. Each round is
composed of $n$ nearest neighbor gates. The first gate in each round is
a one-qubit gate applied to the first qubit. For $i = 2,\ldots, n$, the
$i^{th}$ gate is a two-qubit gate applied to qubits $i-1$ and $i$.
It will be convenient to assume that the first round of gates consists
entirely of identity gates. This will just serve to increase the number
of rounds by one. The value of $R$ is adjusted accordingly.
The construction in \cite{ADKLLR04} also has a sequence of $n$ indentity
gates at the end of each round but we omit that here.
The total number of gates then is $L = nR$.
They observe that any circuit can be transformed to fit this format
by adding additional identity and swap gates.

We now define a Hamiltonian
$H_{prop}$ which will enforce the propogation of the quantum computation.
In this section $H_{final}$ will simply be $H_{prop}$.
The ground state of the $H_{final}$ will be a uniform superposition of
some number of different configurations which we call
{\em templates}. Each template represents
a subspace of $2^n$ possible states and the particular state within
a given template will encode the state of the computation at 
some particular point in time. 
We will think of these templates as a sequence and that the
state of the system changes through time from one template
to the next. 
In fact, the ground state of $H_{final}$ will
be a superposition of these different
snapshots of the computation. 
We start by giving an overview of $H_{prop}$.
In the course of this description, we will describe the states
for each particle and their significance. 

We have a chain of $L+2$ particles, where
$L$ is the number of gates in the quantum circuit.
The particles will be labelled $0$ through $L+1$.
We will think of the particles as organized into $L/n$ contiguous blocks of 
$n$ particles with a single particle on either end (particles $0$ and $L+1$).
The first block of particles are $1$ through $n$, the second block is
$n+1$ through $2n$, etc.
At any point in time, there are $n$ particles whose state
represent the state of the
computation. We will call these the {\em computation particles}.
These are either all contiguous or possibly separated by a
single control particle somewhere in the middle.
In general, $H_{prop}$ will enforce that the computation quibits
shift from the left to the right.
When the computation particles are all aligned within the boundaries of a single block
(i.e. are located in positions $in+1,\ldots,(i+1)n$ for some $i$), 
then the gates are applied for the $i^{th}$  round.

There are 13 possible states for the qubits.

\begin{itemize}
\item {\bf F:} F stands for finished. Particles in this state are to
the left of the computation particles and will no longer change state.
\item {\bf N:} N stands for new. Particles in this state are to the right of the
computation particles and  have not yet been reached.
\item {\bf L:} L stands for left. A particle in this state is used for control and indicates
that the particle in the control state is propogating to the left.
\item ${\bf T_R, T_L:}$ $T_L$ and $T_R$ stand for turn left and right,
respectively. A particle in this state is used for control and indicates
that the control particle will start moving in the opposite direction.
$T_L$ indicates that it will start moving to the left and $T_R$
indicates that it will start moving to the right.
\item ${\bf Q_0, Q_1, B_0, B_1:}$ These states represent qubits. Particles in these states
are computation qubits. The subscript represents the state of a qubit in the computation.
The $B$ states are to the left of the particle in the control state
and the $Q$ states 
are to the right of the particle in the control state.
\item ${\bf R_0, R_1:}$ R stands for right. A particle in this state
is a computation particle as well as a control
particle. The subscript represents the state of a qubit in the computation.
A particle in this state also indicates that the control particle is
moving to the right.
\item {\bf $\bf{G_0, G_1}:$} G stands for gate. A particle in this state
is a computation particle as well as a control
particle. The subscript represents the state of a qubit in the computation.
A particle in this state also indicates that the control particle is
moving to the right and a gate is being applied.
\end{itemize}

We will explain the basic construction of $H_{prop}$ by showing an iteration 
in which the control particle propogates to the right and then to the left.
At the end of the iteration, the computation particles will have
shifted one position to the right. At each step, we indicate the component of
$H_{prop}$ that enforces
the propogation. If $A$, $B$, $X$, $Y$ are possible particle states, we will define 
$$H_{AB \leftrightarrow XY}^i = | AB \rangle \langle AB|_{i,i+1} + | XY \rangle \langle XY|_{i,i+1} 
- | AB \rangle \langle XY|_{i,i+1} - | XY \rangle \langle AB|_{i,i+1} .$$ 
The superscript $i$ indicates that the Hamiltonian is applied
to particles $i$ and $i+1$. This
particular Hamiltonian
enforces the condition that if a state is an eigenstate with eigenvalue $0$,
then for each configuration with $AB$ in positions $i$
and $i+1$ there must be a state of equivalent amplitude with $AB$ replaced by $XY$.

It will be useful in our exposition for each $H_{AB \leftrightarrow XY}^i$
to define two companion tranformations. These correspond to transforming any state
with $AB$ in locations $i$ and $i+1$ by changing the $AB$ to $XY$. This is represented
by $\calf$, where  the $\calf$ stands for propogation in the forward direction.
Similarly $\calb$ will correspond to replacing $XY$ by $AB$ which represents
propogation in the backward direction.
$$\calf_{AB \rightarrow XY}^i = | XY \rangle \langle AB|_{i,i+1}, ~~~
\calb_{AB \leftarrow XY}^i = | AB \rangle \langle XY|_{i,i+1}.$$
These will not be part of the final Hamiltonian but will be helpful
for describing certain states.

For the time being, we will not use that gate states ($G_0$
and $G_1$). This means that the state of the represented quibits will not change
and there is no change of state between $\{Q_0, R_0\}$ and $\{Q_1, R_1 \}$.
For ease of notation, we will omit the subscripts.
For example, $H_{BL \leftrightarrow LQ}$ actually represents 
$H_{B_0 L \leftrightarrow LQ_0 } + H_{B_1 L \leftrightarrow LQ_1 }$.
Similarly the expression $H_{RQ \leftrightarrow BR}$ would be the sum of
four different terms for the possible subscripts on the pair of states.
We will use the term {\em template} to be a representation of
a set of states that all have the same letter value with the
subscripts on the computation bits unspecified.

We start with the particles in the following template:

\begin{center}
\begin{tabular}{|c|c|c|c|c|c|c|c|c|c|c|c|c|}
\hline
F & $\cdots$ & F & $T_R$ & Q & Q & Q & Q & Q & N & N & $\cdots$ & N \\
\hline
\end{tabular}
\end{center}
where $n=5$.
Let particle $i$ be the location of the $T_R$ state.
$H_{T_R Q \leftrightarrow FR}^i$ will result in the following template:
\begin{center}
\begin{tabular}{|c|c|c|c|c|c|c|c|c|c|c|c|c|}
\hline
F & $\cdots$ & F & F & R & Q & Q & Q & Q & N & N & $\cdots$ & N \\
\hline
\end{tabular}
\end{center}
For $j=i+1$ through $i+n-1$ we have the Hamlitonian
$H^j_{RQ \leftrightarrow B R}$ which results in
\begin{center}
\begin{tabular}{|c|c|c|c|c|c|c|c|c|c|c|c|c|}
\hline
F & $\cdots$ & F & F & $B$ & $B$ & $B$ & $B$ & R & N & N & $\cdots$ & N \\
\hline
\end{tabular}
\end{center}
Then $H^{i+n}_{RN \leftrightarrow B T_L}$,
results in 
\begin{center}
\begin{tabular}{|c|c|c|c|c|c|c|c|c|c|c|c|c|}
\hline
F & $\cdots$ & F & F & $B$ & $B$ & $B$ & $B$ & $B$ & $T_L$ & N & $\cdots$ & N \\
\hline
\end{tabular}
\end{center}
The presence of the particle in the $T_L$ (turn left) state indicates that propogation will switch
from moving right to moving left.
$H^{i+n}_{T_L N \leftrightarrow LN}$,
results in 
\begin{center}
\begin{tabular}{|c|c|c|c|c|c|c|c|c|c|c|c|c|}
\hline
F & $\cdots$ & F & F & $B$ & $B$ & $B$ & $B$ & $B$ & L & N & $\cdots$ & N \\
\hline
\end{tabular}
\end{center}
The presence of the particle in state $L$ indicates propogation to the
left via 
$H^j_{B L \leftrightarrow LQ}$
for $j = i+n$ down to $i+1$
until
\begin{center}
\begin{tabular}{|c|c|c|c|c|c|c|c|c|c|c|c|c|}
\hline
F & $\cdots$ & F & F & L & Q & Q & Q & Q & Q & N & $\cdots$ & N \\
\hline
\end{tabular}
\end{center}
Finally, $H^{i+1}_{FL \leftrightarrow FT_R}$ brings us back to the original template
with the computation particles shifted one position to the right:
\begin{center}
\begin{tabular}{|c|c|c|c|c|c|c|c|c|c|c|c|c|}
\hline
F & $\cdots$ & F & F & T & Q & Q & Q & Q & Q & N & $\cdots$ & N \\
\hline
\end{tabular}
\end{center}
There are a total of $2n+3$ configurations in one iteration that shifts the states
one position to the right.

The application of the gates is triggered when the computation qubits
align within the boundaries of a set because we have a slightly different
Hamiltonian at these locations. At the locations $i$ such that $i$ is 
a multiple of $n$,
we have $H^i_{T_R Q \leftrightarrow FG}$ instead of $H^i_{T_R Q \leftrightarrow FR}$. 
In addition,
we apply the qubit gate to $Q_{\oz}$. If the $Q$ is in position $i$,
then this is the $i^{th}$ gate in the computation and we 
denote the $2 \times 2$ unitary matrix associated
with this 1-bit gate as $U_i$. We will express this Hamlitonian 
matrix over
the subspace spanned by
$\{T_R Q_0, T_R Q_1, FG_0, FG_1\}$

\[
H^i_{T_R Q \leftrightarrow FG} =
\left[  
\begin{array}{cc}
           I & -U_i \\
           -U_i^{\dagger} & I \\
          \end{array}
\right],~~~
\calf^i_{T_R Q \rightarrow FG} =
\left[  
\begin{array}{cc}
           0 & U_i \\
           0 & 0 \\
          \end{array}
\right],~~~
\calb^i_{T_R Q \leftarrow FG} =
\left[  
\begin{array}{cc}
           0 & 0 \\
           U_i^{\dagger} & 0 \\
          \end{array}
\right].
\]

Similarly for the 2-qubit gates, we apply the $i^{th}$ gate to 
$GQ$ in locations $i-1$ and $i$ and the result is a state of the
form $B G$. 
We express this Hamiltonian as a matrix over the basis 
$$\{G_0 Q_0, G_0 Q_1, G_1 Q_0, G_1 Q_1, B_0 G_0, B_0 G_1, B_1 G_0, B_1 G_1\},$$
where  the $i^{th}$ gate is expressed by the $4 \times 4$ unitary martix
$U_i$:

\[ 
H^{i-1}_{GQ \leftrightarrow B G} = 
\left[  
\begin{array}{cc}
           I & -U_i \\
           -U_i^{\dagger} & I \\
          \end{array}
\right],~~~
\calf^{i-1}_{GQ \rightarrow B G} = 
\left[  
\begin{array}{cc}
           0 & U_i \\
           0 & 0 \\
          \end{array}
\right],~~~
\calb^{i-1}_{GQ \leftarrow B G} = 
\left[  
\begin{array}{cc}
           0 & 0 \\
           U_i^{\dagger} & 0 \\
          \end{array}
\right].
\]

Finally we need the term that turns the propogation
of the control state when $G$ reaches the rightmost
end of the computation bits. 
For $i \bmod n = 0$, we have
$H_{GN \leftrightarrow B T_L}$ instead of $H_{RN \leftrightarrow B T_L}$.
This just preserves the value of the represented qubit as before.
To summarize, 
$$H_{prop} = \sum_{i=0}^{L} H^i_{prop}.$$
For $i \bmod n \neq 0$
\begin{eqnarray*}
\label{eq:proprule1} 
H^i_{prop}  & =  &
 H^i_{T_R Q \leftrightarrow FR} 
+  H^i_{RQ \leftrightarrow B R} + H^i_{RN \leftrightarrow B T_L} + H^i_{T_L N \leftrightarrow LN}  \\
 &  +  & H^i_{B L \leftrightarrow LQ} + H^i_{GQ \leftrightarrow B G} + H^i_{FL \leftrightarrow FT_R}  
\end{eqnarray*}
For $i \bmod n = 0$, we replace $H^i_{T_R Q \leftrightarrow FR}$ with $H^i_{T_R Q \leftrightarrow FG}$
and $H^i_{RN \leftrightarrow B T_L}$ with $H^i_{GN \leftrightarrow B T_L}$, to get
\begin{eqnarray*}
H^i_{prop} & = &
 H^i_{T_R Q \leftrightarrow FG} 
+  H^i_{RQ \leftrightarrow B R} + H^i_{GN \leftrightarrow B T_L} + H^i_{T_L N \leftrightarrow LN} \\
& + &  H^i_{B L \leftrightarrow LQ} + H^i_{GQ \leftrightarrow B G} + H^i_{FL \leftrightarrow FT_R}  \\
\end{eqnarray*}
Finally, if $i = L$, we remove $H^i_{GN \leftrightarrow B T_L}$ to end the sequence.
We can define $\calf$ by using the corresponding $\calf$ term in place each $H_{prop}$ term.
Similarly we can define $\calb$ by using the corresponding $\calb$ term in place each $H_{prop}$ term.

Now we would like to define the target ground state for $H_{final}$.
In order to define the target ground state, we define a series of templates.
These templates are linearly ordered. The label $t$ for each template ranges from
$0$ through $T = (2n+3)(L-n)+n+1$.
The template is represented by a string of length $L+2$ indicating the
state for each particle, omitting the subscript for any computation bits.
In this section, the templates will always have $n$ computation bits.
Thus, the remaining degrees of freedom represented by the omitted subscripts results
in a subspace of dimension $2^n$.
We say that a term $H_{AB \leftrightarrow XY}$ in
$H_{prop}$ applies in the forward direction to a template if $AB$ appears in string
corresponding to the template.
This means that $\calf_{AB \rightarrow XY}$ has a non-zero result when applied
to a state in the corresponding subspace for that template. The result will be a state
in the template with $AB$ replaced by $XY$. 
Similarly $H_{AB \leftrightarrow XY}$ applies in the backward direction if $XY$ appears in
the template. $\calb_{AB \leftarrow XY}$ applied to a state in that template 
results in a state in the template with $XY$ replaced by $AB$.
Figure \ref{fig:templates} shows
a chart of the templates. The string representing the templates will depend on
the value of $t$ as indicated by the column labelled Conditions.
We also include a column that indicates for each template the set of terms in $H_{prop}$
that apply in the forward direction and a column for the terms that apply in the backwards
direction. 
The condition {\em Boundary} holds for a given term if the pair straddles a boundary
between blocks of $n$ particles. That is, if it applies to particles $p$ and $p+1$
and $p \bmod n = 0$.
For each $t$, let $i = \lfloor t/(2n+3) \rfloor$ and $j = t \bmod (2n+3)$.
\begin{figure}
\label{fig:templates}

$$
\begin{array}{|c|ccc|c|c|}
\hline
 \mbox{Condition} & & \mbox{Template} & &\mbox{Forward} & \mbox{Backward}\\
\hline
\hline
 & {\scriptstyle 0 \cdots i }& {\scriptstyle (i+1) \cdots (m+i) }& {\scriptstyle (n+i+1) \cdots (L+1)} & &\\
\hline
& & & & \scriptstyle{ if~(not ~~boundary)} & \\
 j = 0 & 
{  {F \cdots F} ~T_R }
& 
{  {Q ~~~ \cdots ~~~~ Q}~ } & 
{  N {N \cdots N} } & T_R Q \ra FR & FL \la FT_R \\
& & & & \scriptstyle{ if~(boundary)} & \\
& & & & T_R Q \ra FG & \\
\hline
i \bmod n \neq 0 , ~~
j = 1 
& 
{  
{F \cdots F} ~F } & 
{  
R ~{Q \cdots QQ \cdots Q} 
}
& 
{  N {N \cdots N}  }
& RQ \ra BR & T_R Q \la FR \\
\hline
i \bmod n = 0 , ~~
j = 1 &   
{  
{F \cdots F} ~F 
}& 
{  
G ~{Q \cdots QQ \cdots Q}
}
& 
{  N~ {N \cdots N}
}
& GQ \ra BG & T_R Q \la FG\\
\hline
{\scriptstyle
i \bmod n \neq 0 , ~~~~
1 < j < n }
& 
{  
{F \cdots F} ~F } & 
 ~\underbrace{{  B \cdots B}}_{j}  {  R }
~\underbrace{{  Q \cdots Q}}_{m-j-1}  & 
{   N~ {N \cdots N}  
}
& RQ \ra BR & RQ \la BR\\
\hline
{\scriptstyle
i \bmod n = 0 , ~~~~
1 < j < n }
&  
{  
{F \cdots F} ~F }
& 
  ~\underbrace{{  B \cdots B}}_j {  G } ~\underbrace{{  Q \cdots Q}}_{m-j-1} 
& 
{  N~ {N \cdots N}
}
& GQ \ra BG & GQ \la BG \\
\hline
i \bmod n \neq 0 , ~~
j = n 
&   
{  
{F \cdots F} ~F } & 
{  ~ {B \cdots B} R } & 
{   N~ {N \cdots N} 
}
& RN \ra BT & RQ \la BR\\
\hline

i \bmod n = 0 , ~~
j = n &  
{  
{F \cdots F} ~F } & 
{  B \cdots B  G } & 
{  N~ {N \cdots N} 
}
& GN \ra BT & GQ \la BG\\
\hline
& & & & &  \scriptstyle{ if~(not ~~boundary)} \\
  j = n+1 &   
{  
  {F \cdots F} ~F } & 
{  ~ {B \cdots B} } & 
{   T_L  {N \cdots N} 
  }
  & T_L N \ra LN & RN \la BT_L\\
& & & &  & \scriptstyle{ if~(boundary)} \\
& & & & & GN \la B T_L  \\
\hline
 j = n+2 &  
{  
 {F \cdots F} ~F } & 
{  ~ {B \cdots B}} & 
{  L  {N \cdots N}
 }
 & BL \ra LQ & T_L N \la LN\\
\hline
 n+2 < j < 2n+2   &  
{  
 {F \cdots F} ~F } & 
 ~ \underbrace{ {B \cdots B}}_{m-j} {  L } ~\underbrace{{  Q \cdots Q}}_{j-1} 
& 
{  Q
\underbrace{N \cdots N}_{L-i-m}
}
& BL \ra LQ & BL \la LQ \\
\hline
 j = 2n+2   &  
{  
 {F \cdots F} ~F } & 
{  L }
 \underbrace{{  Q \cdots Q}}_{m-1}~ & 
{  Q {N \cdots N} 
 }
& FL \ra FT_R & LQ \la BL \\
\hline
\end{array}
$$
\caption{Templates}
\end{figure}

\begin{lemma}
\label{lem:state-change}
For each $t \in \{0, \ldots, T-1 \}$, exactly one term in $H_{prop}$
applies to the $t^{th}$ template. Furthermore, when $\calf_{prop}$ is applied to
a state in that template, it results in a state in the next template in the sequence.
For each $t \in \{1, \ldots, T \}$, exactly one term in $H_{prop}$
applies to the $t^{th}$ template in the backward direction. 
Furthermore, when $\calb_{prop}$ is applied to
a state in that template, it results in a state in the previous template in the sequence.
$\calb$ is zero on a state in template $0$ and $\calf$ is $0$ on a state
in template $T$.
\end{lemma}
\proof
The proof consists of verifying that for each template in Figure \ref{fig:templates},
exactly one term applies in the forward direction (and in only one place)
and exactly one term applies in the backward direction (again in only one place).
When the term is applied in the forward direction, the result is the next template
in the sequence and when the term is applied in the backward direction, the result is
the previous template in the sequence.
Template $0$ is $T_R Q^n N^{L-n+1}$ and there are no terms in $H_{prop}$ that
apply in the backward direction.
The $T^{th}$ template is $F^{L-n+1}B^{n-1}GN$. Since $H^L_{GN \lra BT}$ is removed
from $H^L_{prop}$, there is no term in $H_{prop}$ that applies in the forward direction.
\qed

Each template with $n$ computation particles
corresponds to a particular point in the computation to be simulated.
However, the
correspondence is not one to one. A particular point in the computation can
be represented by many templates. The $t^{th}$ template corresponds to the point
in the computation after $g(t)$ gates have been performed, where $g(t)$ is the
number of templates in the sequence from $0$ through $t$
containing a particle in a $G$ state. 
As each new template with a particle in a $G$
state is reached, one more gate is performed.
If $i = \lfloor  t /(2n+3) \rfloor$ and $j = t \bmod (2n+3)$,
then
\[ 
g(t) =
\left\{  
\begin{array}{ll}
           i + \min\{n,j\}  & \mbox{~~~if $i \bmod n = 0$},\\
           n \lceil i/n \rceil & \mbox{~~~if $i \bmod n \neq 0$}\\
          \end{array}
\right.
\]

Define $| \phi_t \rangle$ to be the state in the $t^{th}$ template
such that the superposition of states defined by the $0/1$ subscripts
of the computation bits corresponds to the superposition of
states in the circuit after $g(t)$ gates
have been applied, assuming that the input to the quantum computation
is $|0\rangle$. Thus, $| \phi_0 \rangle = T(Q_0)^n N^{L-n+1}$.

For each $t \in \{0, \ldots, T-1 \}$, $\calf_{prop}| \phi_t \rangle = 
| \phi_{t+1} \rangle$ and 
$\calf| \phi_T \rangle = 0$.
Also, for each $t \in \{1, \ldots, T \}$, $\calb | \phi_t \rangle = 
| \phi_{t-1} \rangle$ and 
$\calb| \phi_0 \rangle = 0$.
Now define $\cals$ to be the subspace spanned by the states
$|\phi_t \rangle$ for all $t \in \{0, \ldots, T \}$.
These states form an orthonormal basis of $\cals$.
$\cals$ is closed under $H_{prop}$.
The matrix representation of $H_{prop}$ in the $|\phi \rangle$ basis
is exactly $P_{T+1}$.
Recall the definition of $P_r$ from Section \ref{sec:SpectralGaps}. 
The unique ground state of 
$H_{final}$ ( $= H_{prop}$) is
$$| \phi_{final} \rangle = \frac{1}{ \sqrt{T+1 }} \sum_{t=0}^{T}  |\phi_{t} \rangle.$$

Now we need to define $H_{init}$:
$$H_{init} = (I - |T_R \rangle \langle T_R | )_0.$$
Note however, that while $H_{init}$ in its current form can distinguish between
$|\phi_0 \rangle$ and the other $|\phi \rangle$'s, it does not ensure that the
initial configuration starts out in $\cals$. 
To address this problem, we add an additional state $S$ (for Start) and some
extra constraints. The initial configuration will be
$T_R S^n N^{L-n+1}$.
We will add a constraint to $H_{init}$ that enforces the condition that if
particle $0$ is in state $T_R$, then particle $1$ is in state $S$.
This is achieved by adding in $|T_R X \rangle \langle T_R X |_0$,
where we sum over all states $X$ such  that $X \neq S$.
Similarly for $i=1$ through $n-1$, we add the constraint that if
particle $i$ is in state $S$, then $i+1$ is also in state $S$.
We also add the constraint that if particle $n$ is in state $S$, then
particle $n+1$ is in state $N$. Finally for all $i=n+1$ through $L$,
we add the constraint that if particle $i$ is in state $N$, then particle
$i+1$ is in state $N$.
All of  these constraints are added into $H_{init}$.

We need to alter $H_{prop}$ slightly in order to work with the $S$
states. In $H^0_{prop}$, we replace $H^0_{T_R Q \leftrightarrow FG}$
with $H^0_{T_R S \leftrightarrow F G_0}$. (Recall that the input is
all $0$'s). Then for $i=1$ through $n-1$, we replace
$H^0_{G Q \leftrightarrow BG}$ with $H^0_{G S \leftrightarrow B_0 G}$.
Templates $1$ through $n$ are now of the form
$F (B_0)^j G_0 S^{n-j-1} N^{L-n+1}$.
Lemma \ref{lem:state-change}
still holds. Furthermore, $H_{init}$ ensures that
there is only one eigenstate with eigenvalue $0$ and it
is our desired initial configuration.

\begin{lemma}
\label{lem:init}
The matrix representation of $H_{init}$ restricted to $\cals$
and expressed in the basis of $| \phi_{t} \rangle$'s
is a diagonal matrix. The first diagonal entry is $0$ and the
others are $1$.
\end{lemma}
\proof
All templates  start with $F$ except the first one 
which starts with $T$. All the templates satisfy the
other conditions in $H_{init}$.
\qed

To establish the spectral gap of $H_{\cals_0}(s)$, we simply
invoke Lemma \ref{lem:sgap} which says that 
the spectral gap of the restriction of $H(s)$ to $\cals_0$
satisfies $\Delta(H_{\cals}(s)) = \Omega ( (Ln)^{-2})$.

Using the Adiabatic Theorem, we get that
the running time of the algorithm will be $O(||H(s)|| \epsilon^{-\delta} 
(Ln)^{4+2 \delta} )$. Note that $||H(s)||$ is $O(L)$ which gives an overall
running time of $O(\epsilon^{-\delta} L (Ln)^{4+2 \delta} )$.
However, this only produces a final state that is $\epsilon$ close
to $ | \phi_{final} \rangle$. This is a superposition of the
$| \phi_t \rangle$'s and only $| \phi_{T} \rangle$
encodes the desired output.
This can be corrected by adding $L/\epsilon$ identity gates to the
computation so that only a fraction of $\epsilon$  of the 
$| \phi_{i,j} \rangle$'s will encode partial points in the computation.
This makes the final running time
$O(\epsilon^{-(5+3\delta)} L (Ln)^{4+2 \delta} )$.

In order to reduce the number of states from 14 to 10, 
we observe that we can identify pairs $(N,F)$,
$(T_R, T_L)$, $(B_0,Q_0)$ and $(B_1,Q_1)$ and Lemma \ref{lem:state-change}
still holds.

\section{8-local Hamiltonian on a Chain is QMA-complete}

We now turn to the problem of the promise local Hamiltonian.
We will continue to work with $13$-state particles as in the previous section.
(We will not use  the $S$ state added at the end of the previous section).
The ground state will now encode the computation performed by a
quantum verifier $V$ which works on input $|x \xi \rangle$
for input $x$ and witness $\xi$.
The total length of the input is $n$ qubits.
We will construct a Hamiltonian such that if there exists a witness
$\xi$ such that $V$ accepts with probability at least $1 - \epsilon$
on input $|0 \xi \rangle$, then the lowest eigenvalue of $H$ will be
less than $\epsilon$.
However, if for every $\xi$, $V$ accepts with probability at most
$\epsilon$, then the lowest eigenvalue will be at least 1 over a
polynomial in $n$ and $L$.

For this problem, we need to show that there is a large spectral gap
over the entire space of the particles, not just when restricted
to a particular subspace. 
The Hamiltonian $H$ will consist of five  components:
$$H = H_{prop} + H_{init} + H_{out} + H_{valid}+ H_{legal}.$$
$H_{prop}$ will the same as in the previous section (with the stipulation
that we do not use the changes made at the end of the
previous section to incorporate the additional state
$S$). 

We will use the term {\em template} here to refer to any $L+2$-character
string over the alphabet $$\{ F, N, B, Q, L, R, G, T_R, T_L \}.$$
As before, each template will designate a subspace according to the
unspecified $0/1$ subscripts for the computation states.
This subspace will have dimension $2^m$ if the template has $m$
computation bits.

\begin{figure}
\label{fig:valid}
\centerline{\psfig{figure=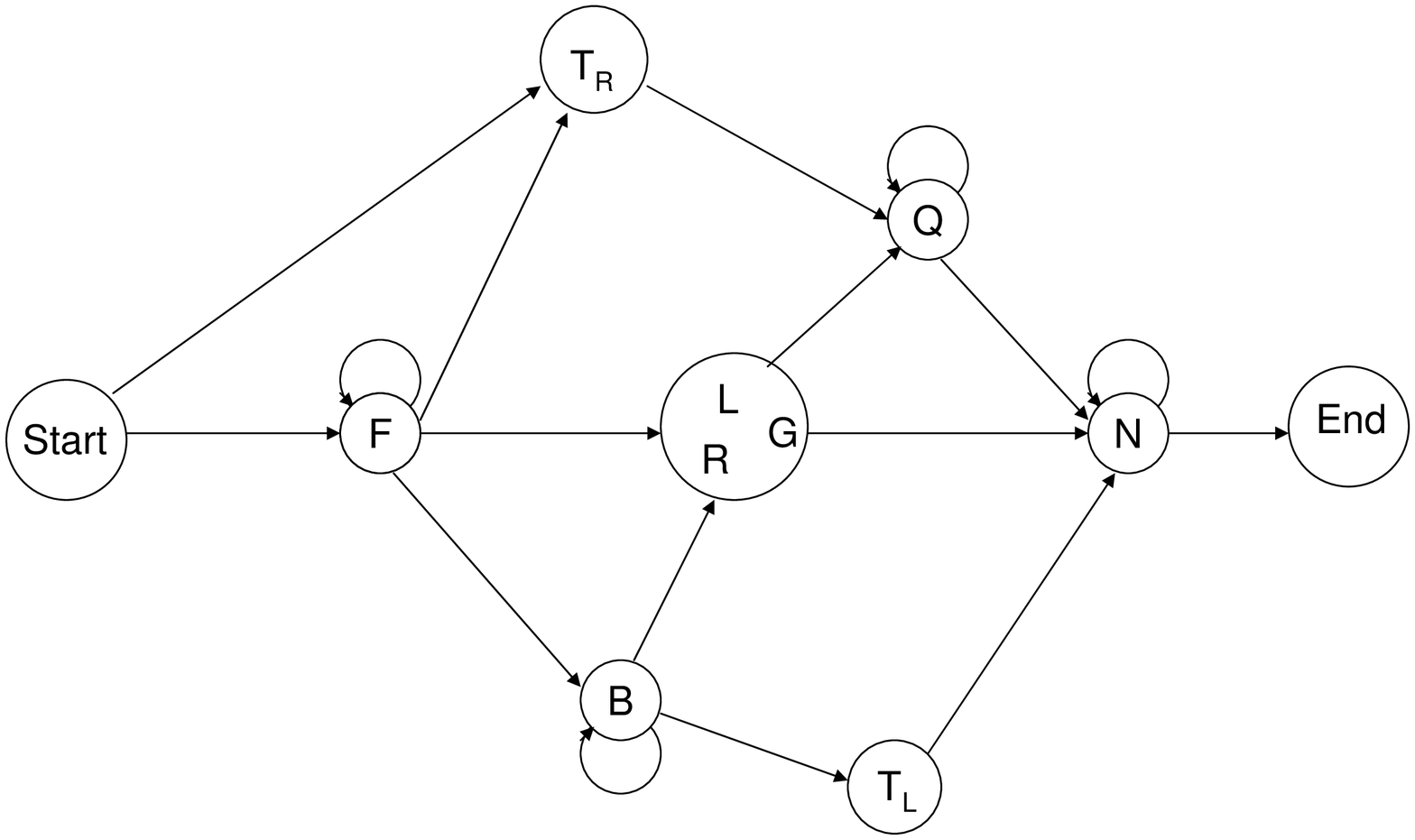,width=5.0in}}
\caption{Graph indicating set of valid templates.}
\end{figure}

We will use various hamiltonians to enforce that only certain templates
will be valid or legal.
We start with a set of consraints enforced in
$H_{valid}$. We will enforce through $H_{valid}$
that a template must have a form which is a string of
length $L+2$ and is specified by one of the three regular expressions:
$$F^+ B^* (R + G + L) Q^* N^+,~~~
F^* T_L Q^+ N^+,  ~~~F^+ B^+ T_R N.$$
Another way to denote these constraints is that a string
corresponding to a valid template must be a path in the
graph in Figure \ref{fig:valid} from the Start node
to the End node consisting of $L+2$ internal nodes.
$R$, $L$, $G$ are grouped together for clarity.
A path through this node can use either $R$, $L$ or $G$.

These constaints are enforced by having a set of allowed
pairs where an edge in the 
graph corresponds to an allowed pair.
We will need some additional constraints to enforce
that the first character must be $F$ or $T_R$ and that the last characers
must be $N$.
For ease of notation, we will omit
the subscripts. Therefore, the pair $RQ$ represents four possible
pairs for the four possible combinations of subscripts. 
When summing over a set, summing over all possible combinations of subscripts
will be assumed.
\begin{eqnarray*}
S & = & \{ FF, FB, BB, FR, FL, FG, BR, BL, BG, RQ, GQ, LQ, \\
    & & QQ, QN, NN, RN, GN, LN, FT_R, T_R Q, BT_L, T_L N\}\\
\end{eqnarray*}
Another way to express the set $S$ is that it is the set below, where the characer $X$
can be either $R$, $L$, or $G$:
\begin{eqnarray*}
S & = & \{ FF, FB, BB, FX, BX, XQ, QQ, QN, NN, XN, FT_R, T_R Q, BT_L, T_L N\}\\
\end{eqnarray*}
There is a one-to-one correspondence between the above pairs and the
edges in Figure \ref{fig:valid}.

For any $i$ from $0$ to $T-1$, we have
$$ H^{i}_{valid} =  I - \sum_{\alpha \beta \in S} | \alpha \beta \rangle \langle \alpha \beta |_{i,i+1}
. $$
We then sum these together and add some additional constraints at the beginning and end of the chain.
$$H_{valid} = (I - |F\rangle \langle F| - |T_R \rangle \langle T_R|)_0
+ (I - | N \rangle \langle N | )_{L+1} + \sum_{i=0}^L H^{i}_{legal} .$$
Note that $H_{valid}$ is a diagonal matrix with non-negative integers
along the diagonal.

\begin{lemma}
Consider a state $| \phi \rangle$ contained in a subspace 
specified by a template.  $H_{valid} | \phi \rangle  = \lambda | \phi \rangle $
for some non-negative integer $\lambda$.
If $\lambda = 0$,
then the template must have a form specified by one of the following
three regular expressions:
$$F^+ B^* (R + G + L) Q^* N^+,~~~
F^* T_L Q^+ N^+,  ~~~F^+ B^+ T_R N.$$
\end{lemma} 

\proof
If $H_{valid} | \phi \rangle = 0$,
then every consecutive pair of characters in the 
the template containing $ |\phi \rangle$ must be an allowable pair.
Furthermore, the first character must be $F$ or $T_R$ and the last
character must be $N$.
The graph in Figure \ref{fig:valid}
has every allowable pair
labelled as a directed edge. Furthermore, the graph
enforces that any path from {\em Start} to
{\em End} must begin with an $F$ or a $T_R$ and
must end with an $N$.
Therefore, a valid template must correspond to a path that
starts at the {\em Start} node, ends at the {\em End}
node and traverses $L+2$ intermediate nodes.
The length of the template is enforced by the physical length
of the chain of particles.
The set of all such paths correspond to the three
regular expressions above with the constraint that the length must be
$L+2$.
\qed

We will label each valid template with a pair $(m,t)$.
$m$ will designate the number of computation bits
in a template. We describe here
how to determine $t$ for a particular $m$.
We describe how to determine two inetegers $i$ and
$j$ and then let $t = i(2m+3) + j$. The labelling will have four
distinct cases, depending on the form of the template:
$$
\begin{array}{|c|c|c|}
\hline
F^k T_L Q^m N^{L-m-k+1} & i \leftarrow k & j \leftarrow 0 \\
\hline
F^{k+1} B^{l} (G + R) Q^{m-l-1} N^{L-m-k+1} & i \leftarrow k & j \leftarrow l+1 \\
\hline
F^{k+1}  B^m T_R N^{L-m-k+1} & i \leftarrow k & j \leftarrow m+1 \\
\hline
F^{k+1} B^{m-l} L Q^{l} N^{L-m-k} & i \leftarrow k & j \leftarrow m+2+l \\
\hline
\end{array}
$$

Conversely, Figure \ref{fig:gen-templates} shows for a given pair
$(m,t)$ the form of the templates corresponding to that pair,
where $i = \lfloor t/(2m+3) \rfloor$ and $j = t \bmod (2m+3)$.
Note that for each pair, there is exactly one template unless
$1 \le j \le m$ in which case there are two templates, depending
on whether the particle in the control state is in a $G$ or an
$R$ state.
Define $T_m = (2m+3)(L-m)+m+1$.
Also note that for a given $m$, and any
$t \in \{0, \ldots, T_m \}$, $(m,t)$ corresponds to a valid template,
but having $t > T_m$ implies that the last particle
will not be in state $N$, which makes the template invalid.
If a template is labelled $(m,0)$, then we will call it an 
{\em initial} template.
If it is labelled $(m, T_m)$, then we will call it
a {\em final} template.

Figure \ref{fig:gen-templates} also shows
for each template which terms in $H_{prop}$
apply in the forward direction and which apply in the backward direction.
The condition $(boundary)$ indicates whether a pair straddles two
blocks of particles. That is, if a pair of particles are located in positions
$i$ and $i+1$ and $i \bmod n = 0$.

\begin{figure}
\label{fig:gen-templates}
$$
\begin{array}{|c|ccc|c|c|}
\hline
 \mbox{Condition} & & \mbox{Template} & &\mbox{Forward} & \mbox{Backward}\\
\hline
\hline
 & {\scriptstyle 0 \cdots i }& {\scriptstyle (i+1) \cdots (m+i) }& {\scriptstyle (m+i+1) \cdots (L+1)} & &\\
\hline
& & & & \scriptstyle{ if~(not ~~boundary)} & \\
 j = 0 & {F \cdots F} ~T_R &{Q ~~~ \cdots ~~~~ Q}~ & N {N \cdots N} & T_R Q \ra FR & FL \la FT_R \\
& & & & \scriptstyle{ if~(boundary)} & \\
& & & & T_R Q \ra FG & \\
\hline
& & & & & \scriptstyle{ if~(not ~~boundary)} \\
j = 1
&  {F \cdots F} ~F  & R ~{Q \cdots QQ \cdots Q} & N {N \cdots N}  & RQ \ra BR & T_R Q \la FR \\
\hline
& & & & & \scriptstyle{ if~(boundary)} \\
j = 1 & {F \cdots F} ~F  & G ~{Q \cdots QQ \cdots Q} & N {N \cdots N}  & RQ \ra BG & T_R Q \la FG \\
\hline
1 < j < n 
&   {F \cdots F} ~F & ~\underbrace{B \cdots B}_{j}  R 
~\underbrace{Q \cdots Q}_{m-j-1}&  N~ {N \cdots N}  & RQ \ra BR & RQ \la BR\\
\hline
1 < j < n 
&   {F \cdots F} ~F & ~\underbrace{B \cdots B}_{j}  G 
~\underbrace{Q \cdots Q}_{m-j-1}&  N~ {N \cdots N}  & GQ \ra BG & GQ \la BG \\
\hline
& & & &  \scriptstyle{ if~(not ~~boundary)} &\\
j = n 
&   {F \cdots F} ~F & ~ {B \cdots B} R &  N~ {N \cdots N}    & RN \ra BT & RQ \la BR\\
\hline
& & & & \scriptstyle{ if~(boundary)} & \\
j = n 
&    {F \cdots F} ~F & B \cdots B  G & N~ {N \cdots N}  & GN \ra BT & GQ \la BG\\
\hline
& & & & &  \scriptstyle{ if~(not ~~boundary)} \\
  j = n+1 &   {F \cdots F} ~F & ~ {B \cdots B} &  T_L  {N \cdots N} & T_L N \ra LN & RN \la BT_L\\
& & & &  & \scriptstyle{ if~(boundary)} \\
& & & & & GN \la B T_L  \\
\hline
 j = n+2 &   {F \cdots F} ~F & ~ {B \cdots B} & L  {N \cdots N} & BL \ra LQ & T_L N \la LN\\
\hline
 n+2 < j < 2n+2   &   {F \cdots F} ~F & ~ \underbrace{B \cdots B}_{m-j} L ~\underbrace{Q \cdots Q}_{j-1}
& Q
\underbrace{N \cdots N}_{L-i-m} & BL \ra LQ & BL \la LQ \\
\hline
 j = 2n+2   &   {F \cdots F} ~F  & L \underbrace{Q \cdots Q}_{m-1}~ & Q {N \cdots N} 
& FL \ra FT_R & LQ \la BL \\
\hline
\end{array}
$$
\caption{Generalized Templates}
\end{figure}

\begin{lemma}
\label{lem:state-change2}
\label{lem:valid}
Consider a valid template $\calt$ labelled with $(m,t)$.
At most one term in $H_{prop}$
applies to $\calt$ in the forward direction.
 Furthermore, when $\calf$ is applied to
a state in that template, the result is $0$ or a state in an $(m,t+1)$-template.
At most one term in $H_{prop}$
applies to $\calt$ in the backward direction.
 Furthermore, when $\calb$ is applied to
a state in that template, the result is $0$ or a state in an $(m,t-1)$-template.
\end{lemma}
\proof
The proof consists of verifying that for each template in Figure \ref{fig:gen-templates},
at most one term applies in the forward direction (and in only one place)
and exactly one term applies in the backward direction (again in only one place).
When the term is applied in the forward direction, the result is the next template
in the sequence and when the term is applied in the backward direction, the result is
the previous template in the sequence.
\qed

\begin{lemma}
\label{lem:end}
Consider a state $| \phi \rangle$ contained in the subspace corresponding to a template $\calt$.
If $\calt$ is an initial template, then $\calb | \phi \rangle = 0$.
If $\calt$ is a final template, then $\calf | \phi \rangle = 0$.
\end{lemma}
\proof
There is no term in $H_{prop}$ that applies to an inital template
in  the backward direction. Furthermore, 
since $H^L_{GN \lra BT}$ is removed
from $H^L_{prop}$, there is no term in $H_{prop}$ that applies to a
final template in the forward direction.
\qed

We define another Hamiltonion $H_{legal}$ which penalizes any 
template for which there is no term in $H_{prop}$ that applies in the
backward direction or for which there is no term that applies in the forward direction
(unless it is an inital or final template, respectively).
Using the table in Figure
\ref{fig:gen-templates}, we want to forbid pairs $RN$ and $FR$ from crossing a boundary.
We will also forbid pairs $GQ$ and $BG$ from crossing a boundary.
In addition, we want to forbid pairs $GN$ and $FG$ unless they cross a boundary.

For any $i$ such that $i \bmod n = 0$, we have
$$ H^{i}_{legal} =   | RN \rangle \langle RN |_i + | FR \rangle \langle FR |_i
+ | GQ \rangle \langle GQ |_i + | BG \rangle \langle BG |_i.$$
For any $i$ such that $i \bmod n \neq 0$, we have
$$ H^{i}_{legal} = | GN \rangle \langle GN |_i + | FG \rangle \langle FG |_i.$$
As usual, summing over all the subscripts of the computation states is assumed.
Finally, we sum these together 
$$H_{legal} =  \sum_{i=0}^L H^{i}_{legal} .$$
We say that a template is legal if any state $| \phi \rangle$
in that template has $H_{legal} | \phi \rangle = 0$.
Otherwise, $H_{legal} | \phi \rangle \ge 1$ and we say the template is
{\em illegal}.

\begin{lemma}
\label{lem:legal}
Consider a template $\calt$ that is both valid and legal.
If $\calt$ is not a final template, then there is a term in $H_{prop}$
that applies to $\calt$ in the forward direction.
Similarly, if $\calt$ is not an initial template, there is a
term in $H_{prop}$ that applies to $\calt$ in the backward
direction.
\end{lemma}
\proof
The proof consists of the obervation that any template
in Figure \ref{fig:gen-templates} for which there is no
term in $H_{prop}$ that applies in the forward direction is made
illegal by $H_{legal}$.Similarly, any template for which there is no
term in $H_{prop}$ that applies in the backward direction is made
illegal by $H_{legal}$.
\qed

We can think of all the valid templates as nodes in a directed graph.
There is a directed edge from templates $\calt$ to $\calt'$ if
applying $\calf$ to some state in $\calt$ results in a state in
$\calt'$. (By definition then, applying $\calb$ to a state in $\calt'$
results in a state in $\calt$).
We will call this graph the {\em template graph} and will refer
to nodes and templates interchangeably.
Lemmas \ref{lem:valid}, \ref{lem:end} and \ref{lem:legal}
imply that this graph consists of a set of disjoint chains.
All the nodes in a chain correspond to templates with the same
number of computation particles. Furthermore, the starting node
in a maximal chain is either an initial template or an illegal one.
The last node in a maximal chain is either a final template or an illegal one.

\begin{lemma}
There is exactly one chain in the template graph that contains no
illegal nodes. Furthermore, templates in this chain have $n$ computation
particles.
\end{lemma}
\proof
Consider a maximal chain with no illegal nodes. This chain must
begin with an initial template and end with a final one.
The initial template has the form $T_R Q^m N^{L-m+1}$ for some $m$.
By the forward rule $T_R Q \rightarrow FG$, the next node in the chain
is $F G Q^{m-1} N^{L-m+1}$. If $m < n$, then $m-1$ applications for
the forward rule $GQ \rightarrow BG$ will result in
$F B^{m-1} G N^{L-m-1}$. Since $m < n$, this will result in a 
pair $GN$ in locations $m$ and $m+1$
which does not straddle a block boundary. This is made illegal
in $H_{legal}$.
If $m > n$, then $n-1$ applications of the forward rule $GQ \rightarrow BG$
to $F G Q^{m-1} N^{L-m+1}$ will result in 
$F B^{n-1} G Q^{m-n} N^{L-m+1}$. This will result in a pair
$GQ$ in locations $n$ and $n+1$ which is also disallowed in $H_{legal}$.

Finally, if $m=n$, there is exactly one chain from an initial template
to a final one (because there is exactly one initial template).
The application of the forward rules do not result in any illegal tempaltes.
\qed

It will be convenient at this point to define the remaining two terms in
$H$.  The input to the quantum verifier will be $n$ qubits.
$n_1$ qubits will be ancillary qubits that are initialized to $0$
and $n_2$ qubits will be the witness $\xi$.
$n_1 + n_2 = n$. We force $x=0$ with the
following Hamiltonian:
$$ H_{input} = \sum_{i=1}^{n_1}
| Q_1 \rangle \langle Q_1 |_i + | R_1 \rangle \langle R_1 |_i + 
| G_1 \rangle \langle G_1 |_i
+ | B_1 \rangle \langle B_1 |_i.$$
We will assume that the output will be present in the rightmost
qubit in of the computation. Therefore, we will have
$H_{out}$ defined to be $ | G_0 \rangle \langle G_0 |_{L}$.
Observe that $H_{input}$ and $H_{out}$ are both closed over the subspace
defined by each template.

A maximal chain in the template graph defines a subspace which is just the
subspace spanned by all the subspaces defined by the templates along the chain.
$H_{prop}$ is closed over the subspace defined by any maximal chain in the
template graph. All the other terms in $H$ are closed over
the subspace defined by each template.

Define $\cals_{legal}$ to be the subspace defined by 
the unique chain containing only legal nodes.
Let $\cals_{legal}^{\perp}$ be the orthogonal subspace to  $\cals_{legal}$.
Since $H$ is closed under $\cals_{legal}$, it is also closed
under $\cals_{legal}^{\perp}$
and any eigenvector of $H$ must be
contained in $\cals_{legal}$ or $\cals_{legal}^{\perp}$.

\begin{lemma}
Any eigenvector of $H$ in $\cals_{legal}^{\perp}$
will have an eigenvalue of at least $\Omega(1/L^4)$.
\end{lemma}

\proof
Define $\cals_{valid}$ to be the subspace spanned by all the valid
templates. Since $H$ is closed over $\cals_{valid}$, it is also closed
under the orthogonal space $\cals_{valid}^{\perp}$.
Any state eigenstate of $H_{valid}$ in $\cals_{valid}^{\perp}$
will have an eigenvalue of at least $1$. Since the remaining terms 
in $H$ are positive semi-defininte, any eigenstate of $H$ in
$\cals_{valid}^{\perp}$ will also have an eigenvalue of at least $1$.

Now we can carve up $H_{valid}$ into the subspaces defined by the maximal
chains and $H$ is closed on each such subspace. We  
focus on an arbitrary such maximal chain (except  the one
containing only legal nodes) and the subspace it defines.
 The chain goes from an
 $(m,t_1)$ template to an $(m,t_2)$-template for some $m$, $t_1$ and $t_2$.
 To specify a state in the first template, we specify an $m$-bit $x$
 string which will determine the subscripts of particles in the computation states.
 We call this state $| \phi_{x,m,t_1} \rangle$. 
 The set of these states for all $x$ forms a basis of the first template.
 Since $\calf$ is unitary,
 when we apply $\calf$ to all the $| \phi \rangle$'s, we get a basis of
 the next template in the chain. 
 Applying $\calf$ $i-1$ times gives a basis
 of the $i^{th}$ template in the chain.
 We will focus on a sequence
 $| \phi_{x,m,t_1} \rangle, \ldots, | \phi_{x,m,t_2} \rangle$,
 where $\calf^i | \phi_{x,m,t_1} \rangle = | \phi_{x,m,t_1+i} \rangle$.
 The subspace spanned by these states is closed under $H_{prop}+H_{legal}$.
 Furthermore $H_{prop}$ when restricted to this subspace and expressed
 in the basis of $\phi$'s is $P_r$, for $r = t_2 - t_1 +1$.
 $H_{legal}$ when expressed in this basis is diagonal with non-negative
 integer entries. Furthermore, we know that there is
 at least one positive entry on the diagonal because the
 chain has at least one illegal node.
 By Lemma \ref{lem:gap}, we know that the lowest eigenvalue of any
 eigenstate in the subspace defined by the chain must have an eigenvalue
 that is at least $\Omega(1 / (T_n)^3 )$.
 Since the remaining terms in $H$ are all positive semi-defininte,
 the lower bound holds for $H$ as well.
 \qed

We will prove the following theorem:

\begin{theorem}
\label{th:complete}
If the circuit $V$ accepts with probability at least $1 - \epsilon$ on
some input $| 0 \xi \rangle$, then $H$ has an eigenvalue smaller than $\epsilon$.
If the circuit $V$ accepts with probability less than $\epsilon$ on all inputs
$| 0 \xi \rangle$, then all eigenvalues of $H$ are larger than $1$
over a polynomial in $n$ and $L$.
\end{theorem}

The input to $V$ consists of $n_1$ auxiliary qubits and $n_2$ witness qubits, where
$n_1 + n_2 = n$.
Together a string $x$ of $n_1$ bits and $\xi$ of $n_2$ bits defines an input to
$V$.
Let $|\phi_{x, \xi, 0} \rangle$ be the state in template $(n,0)$ such that
the input bits are set to $x$ and $\xi$.
Let $|\phi_{x, \xi, t} \rangle = \calf^t |\phi_{x, \xi, 0} \rangle$ for
$t \in \{1, \ldots, T_n \}$.
The space $\cals_{legal}$ is spanned by these $|\phi \rangle$'s.

We define
$$|\nu_{x,\xi} \rangle = \frac{1}{ \sqrt{T_n + 1} } \sum_{i=0}^{T_n}
|\phi_{x,\xi,t} \rangle  .$$
The following lemma established Theorem \ref{th:complete}
in one direction.

\begin{lemma}
If there is a $\xi$ such that $V$ accepts with probability at least $1 - \epsilon$ on
some input $| 0 \xi \rangle$, 
then the smallest eigenvalue of $H$ is at most $\epsilon$.
\end{lemma}
\proof
Let $|\nu \rangle = | \nu_{0,\xi} \rangle$.
$$ \langle \nu | H_{prop} |\nu \rangle =
\langle \nu | H_{legal} |\nu \rangle =
\langle \nu | H_{valid} |\nu \rangle =
\langle \nu | H_{in} |\nu \rangle =0.$$
For $t \in \{0, \ldots, T_n -1 \}$, particle $L$ is not in a $G$
state, so the conditions of $H_{out}$ are satisfied.
For $t = T_n$, if the system is in state
$\|phi_{0,\xi,t} \rangle$ and the $L^{th}$ particle is measured,
the probability that the outcome is $G_0$ is at most $\epsilon$.
Therefore $\langle \nu | H_{out} |\nu \rangle \le \epsilon$.
\qed

The final step is the following lemma:

\begin{lemma}
If for all $\xi$,
$V$ accepts with probability at most  $\epsilon \le 1/2$ on
input $| 0 \xi \rangle$, 
then the smallest eigenvalue of $H$ in $\cals_{legal}$
is at least  an
inverse polynomial in  $L$.
\end{lemma}

\proof
We will use Lemma \ref{lem:geom}.
With $H_1 = H_{prop}$ and $H_2 = H_{in} + H_{out}$.
All of $\cals_{legal}$ is zero on the other two terms,
$H_{legal}$ and $H_{valid}$.
The ground space of $H_{prop}$ is spanned by
the $|\nu_{x,\xi} \rangle$'s, and the smallest non-zero eigenvalue
is $\Omega(1 / (T_n)^2)$.
Similarly for $H_2$, the ground space can be expressed
as any linear combination of states that have $G_1$ in
particle $L$ and do not have $Q_1$, $R_1$, $G_1$ or $B_1$ in
particles $1$ through $n$.
Any non-zero eigenvalue is at least $1$.

Let $P_2$ be the projection onto the ground space of $H_2$.
The cosine of the angle between any state $|\phi \rangle$
and the ground state of $H_2$ is just
$\langle \phi | P_2 | \phi \rangle$.
We will show that for any $|\nu_{x,\xi} \rangle$, the
cosine of the angle between $|\nu_{x,\xi} \rangle$ and the ground space of $H_2$
$$\langle \nu_{x,\xi} | P_2 | \nu_{x,\xi}\rangle \le 1 - \frac {1 - \epsilon}  {T_n + 1}.$$
Thus, for small enough $\epsilon$, 
the sine of the angle between the ground space of $H_1$ and the
ground space of $H_2$ is at least $\Omega(1/T_n)$.

The amplitude of $ | \phi_{x,\xi,0} \rangle$ in $| \nu_{x,\xi} \rangle$
is at least $1 / \sqrt{T_n +1}$. If $x \neq 0$, this state has a $Q_1$ somewhere
in the first $n$ particles and $\langle \phi_{x,\xi,0} | P_2 | \phi_{x,\xi,0} \rangle =0$.
Therefore, 
$\langle \nu_{x,\xi} | P_2 | \nu_{x,\xi}\rangle \le 1 - \frac 1 {T_n + 1} $.
If $x=0$, then we know that for all $\xi$,
$$\langle \Phi_{0,\xi,T_n} | (| G_1 \rangle \langle G_1 |)_L |\Phi_{0,\xi,T_n} \rangle \le \epsilon,$$
because this is the probability that the outcome on input $\xi$ is $1$.
This means that 
$$\langle \nu_{0,\xi} | P_2 | \nu_{0,\xi}\rangle \le 1 - \frac {1 - \epsilon}  T .$$
\qed

\bibliographystyle{plain}
\bibliography{quantum}

\end{document}